\documentclass[twocolumn,showpacs,preprintnumbers,amsmath,amssymb]{revtex4}

\usepackage{graphicx}
\usepackage{dcolumn}
\usepackage{bm}

\begin{document}


\title{Irrotational Momentum Fluctuations Conditioning the Quantum
Nature of Physical Processes}

\author{Stephan I. Tzenov}
\email{tzenov@sa.infn.it}

\affiliation{Dipartimento di Fisica "E.R. Caianiello",
Universit\'a degli Studi di Salerno and INFN Sezione di Napoli
\\Gruppo Collegato di Salerno, Via S. Allende, I-84081 Baronissi (SA), Italy}%


\date{\today}

\begin{abstract}
Starting from a simple classical framework and employing some
stochastic concepts, the basic ingredients of the quantum
formalism are recovered. It has been shown that the traditional
axiomatic structure of quantum mechanics can be rebuilt, so that
the quantum mechanical framework resembles to a large extent that
of the classical statistical mechanics and hydrodynamics. The main
assumption used here is the existence of a random irrotational
component in the classical momentum. Various basic elements of the
quantum formalism (calculation of expectation values, the
Heisenberg uncertainty principle, the correspondence principle)
are recovered by applying traditional techniques, borrowed from
classical statistical mechanics.
\end{abstract}

\maketitle

\renewcommand{\theequation}{\thesection.\arabic{equation}}

\setcounter{equation}{0}

\section{\label{sec:intro}Introduction}

One of the most intriguing achievement of the 20-th century
physics is the foundation of quantum mechanics and its basic tool
the Schr\"{o}dinger equation. It is one of the most studied
equation in contemporary physics both from mathematical point of
view, as well as from the perspective of the enormous number of
its important applications. Various links have been proposed
between classical and quantum picture, in order to overcome the
main difficulties due to the difference of formalisms, and to
better understand their interplay and possible connection.

A formal approach is represented by the Wigner-Weyl transformation
\cite{Weyl,Wigner}, based on the reformulation of quantum
mechanics into phase space by means of a continuous map. Although
having some drawbacks, such as the fact that the Wigner
distribution function is not strictly positive, this approach has
proved an important tool in many areas of quantum physics and
chemistry. In particular, the semi-classical limit of quantum
mechanics can be recovered by using the Wigner-Weyl
transformation.

While the Wigner-Weyl approach can be considered as an extension
of quantum mechanics in a classical domain (phase space), the
Koopman-von Neumann approach \cite{Koopman,Neumann1,Neumann2} is a
reformulation of classical mechanics into the Hilbert-space
language. This quantum-like theory, although being always
classical has the advantage that it can be directly compared to
quantum mechanics, at least in terms of the underlying formalism.

The two guidelines of development mentioned above are actually the
two sides of the same subject, namely to find a formal
mathematical language capable to handle both classical and quantum
mechanics. There exists a third attempt fundamentally different
compared to those, which is known as the Bohm-F\'{e}nyes-Nelson
approach \cite{Bohm,Vigier,Fenyes,Nelson1,Nelson2,Nelson3}. This
approach is centered on the physical footings and interpretation
of quantum mechanics. The basic idea is that quantum phenomena are
conditioned by stochastic effects, which take place in a classical
framework, so that the notion of equations of motion and hence the
notion of trajectory (although a random one) remains valid. The
Planck's constant $\hbar$ plays now the role of a measure of the
strength of the stochastic effects. This is an interesting point
of view exhibiting a number of remarkable properties and links to
the theory of deformation quantization. Many articles
\cite{Polavieja,Leavens,Prata1,Prata2} are devoted to the study of
the relation between the phase space representation of quantum
states (quasi-distributions) in the deformation quantization
method, and the Bohm distributions in phase space in the framework
of the Bohm-F\'{e}nyes-Nelson approach.

Recently, it has been shown \cite{Hall} that an exact uncertainty
principle can be formulated, which provides the key argument in
the transition from the dynamical description of a classical
ensemble to that of a quantum ensemble. Another interesting
derivation of the equations of nonrelativistic quantum mechanics
is based on the use of the principle of minimum Fisher information
\cite{Reginatto}.

The point of view pursued in the present paper has been formulated
previously in a different context \cite{Tzenov}. Its cornerstone
is the assumption that the particle velocity, being an
infinitesimal quantity (derivative of the position with respect to
time), is random and consists of a mean part and an irrotational
fluctuation. The mean part is the actual classical velocity
(momentum), while the fluctuation represents a measure of the
uncertainty with which it can be determined. Since the fluctuation
is irrotational, it does not affect the averaged Hamilton
equations of motion, however it yields an additional second-order
correlation term in a picture, where an ideal particle
localization is not valid. Moreover, as it will become clear in
the sequel, the strength of the momentum fluctuation depends
functionally on the measure of particle delocalization, namely on
the density distribution $\varrho {\left( {\bf x}; t \right)}$.

The paper is organized as follows. In the next Section a brief
parallel between the Hamiltonian and Liouvillean description is
outlined. In Section III, a mapping of the classical kinetic
balance equations onto a Schr\"{o}dinger equation is derived. The
second-order correlation tensor of the random, irrotational
velocity field is derived in Section IV. Further, it is shown that
the hydrodynamic equations thus obtained, are in fact the Madelung
equations \cite{Madelung}, known to be formally equivalent to the
Schr\"{o}dinger equation. Section V presents a further elaboration
aimed to recover some basic rules of the quantum formalism.
Finally, Section VI is devoted to discussion and conclusions.

\renewcommand{\theequation}{\thesection.\arabic{equation}}

\setcounter{equation}{0}

\section{Liouvillean Description}

To start with, we consider a $n$-dimensional Hamiltonian system,
whose dynamics is governed by a smooth Hamiltonian function $H
{\left( {\bf x}, {\bf p}; t \right)}$. Here ${\bf x}$ and ${\bf
p}$ are $n$-dimensional vectors. It is well-known that the
description of the dynamical system in terms of equations of
motion (Hamilton's equations) is formally equivalent to the
description in terms of a phase space density specified by a
distribution function $f {\left( {\bf x}, {\bf p}; t \right)}$.
The latter satisfies the Liouville equation
\begin{equation}
{\frac {\partial f} {\partial t}} + {\left\{ f, H \right\}} = 0,
\label{eq:liouville}
\end{equation}
where
\begin{equation}
{\left\{ F, G \right\}} = \sum \limits_{k=1}^n {\left( {\frac
{\partial F} {\partial x_k}} {\frac {\partial G} {\partial p_k}} -
{\frac {\partial G} {\partial x_k}} {\frac {\partial F} {\partial
p_k}} \right)} \nonumber
\end{equation}
\begin{equation}
= {\left( \nabla_x F \right)} \cdot {\left( \nabla_p G \right)} -
{\left( \nabla_x G \right)} \cdot {\left( \nabla_p F \right)},
\label{eq:poisbra}
\end{equation}
is the standard Poisson bracket. This equivalence is however
subtle. Actually, it is a formal mathematical property, which is
evidently incomplete in a physical sense. Note for example that,
unlike the Hamilton's equations of motion
\begin{equation}
{\frac {{\rm d} {\bf x}} {{\rm d} t}} = \nabla_p H, \qquad {\frac
{{\rm d} {\bf p}} {{\rm d} t}} = - \nabla_x H,
\label{eq:hamiltequat}
\end{equation}
where ${\bf x} = {\bf x} (t)$ and ${\bf p} = {\bf p} (t)$, in
equation (\ref{eq:liouville}) ${\bf x}$ and ${\bf p}$ are
independent variables and they do not depend on $t$. Moreover, the
Hamilton equations correspond to the ideal case of a perfectly
localized particle
\begin{equation}
f {\left( {\bf x}, {\bf p}; t \right)} = \delta {\left[ {\bf x} -
{\bf X} (t) \right]} \delta {\left[ {\bf p} - {\bf P} (t)
\right]}, \label{eq:localized}
\end{equation}
where ${\left( {\bf X} (t), {\bf P} (t) \right)}$ represents their
explicit solution. Expression (\ref{eq:localized}) also satisfies
the Liouville equation, however in addition, it admits a solution
of the form (\ref{eq:exacrsolut}). The latter implies that for a
given value of ${\bf x}$, the classical momentum ${\bf p}$ is
uniquely determined according to the expression [see equation
(\ref{eq:eikonal})]
\begin{equation}
{\bf p} = \nabla_x {\cal S} {\left( {\bf x}; t \right)}.
\label{eq:clasmoment}
\end{equation}
In addition, the function ${\cal S} {\left( {\bf x}; t \right)}$
satisfies the Hamilton-Jacobi equation and represents a family of
classical trajectories. Evidently, the density distribution
$\varrho {\left( {\bf x}; t \right)}$ is a new element in general.
It can be regarded as a generalization of the delta-distribution
in the Hamiltonian description, and subsequently as a measure of
particle delocalization.

Let us define the characteristic function
\begin{equation}
{\cal G} {\left( {\bf x}, {\bf s}; t \right)} = \int {\rm d} {\bf
p} f {\left( {\bf x}, {\bf p}; t \right)} \exp {\left( {\frac {i}
{\lambda}} {\bf s} \cdot {\bf p} \right)}, \label{eq:charfunc}
\end{equation}
with a Fourier inverse
\begin{equation}
f {\left( {\bf x}, {\bf p}; t \right)} = {\frac {1} {{\left( 2 \pi
\lambda \right)}^n}} \int {\rm d} {\bf s} {\cal G} {\left( {\bf
x}, {\bf s}; t \right)} \exp {\left( - {\frac {i} {\lambda}} {\bf
s} \cdot {\bf p} \right)}, \label{eq:charfunct}
\end{equation}
where the variable ${\bf s}$ is chosen such that to have the same
dimension as the coordinate ${\bf x}$, and $\lambda$ is a formal
parameter with dimension of action. It is introduced in order to
make the argument under the exponent dimensionless. In fact,
equation (\ref{eq:charfunc}) [or equivalently
(\ref{eq:charfunct})] represents the definition of the Wigner-Weyl
transformation \cite{Wigner}. For the time being, we avoid any
reference to the latter, apart from formal similarity and
coincidence. In Section V, we will establish a firm link between
the characteristic functional and the Wigner-Weyl representation.

Before we proceed further, let us point out a couple of important
features of the characteristic function. It is known that in the
limit ${\bf s} \rightarrow 0$, the characteristic function yields
the moments of the distribution function $f {\left( {\bf x}, {\bf
p}; t \right)}$ after a proper marginalization with respect to the
conjugate momentum variable ${\bf p}$ is performed. For example,
\begin{equation}
\lim \limits_{{\bf s} \rightarrow 0} {\cal G} {\left( {\bf x},
{\bf s}; t \right)} = {\frac {\varrho {\left( {\bf x}; t \right)}}
{m {\cal N}}}, \label{eq:density}
\end{equation}
where $m$ is the particle mass, and
\begin{equation}
{\cal N} = \lim \limits_{N,V \rightarrow \infty} {\frac {N} {V}},
\label{eq:numdensity}
\end{equation}
is the particle number density in the thermodynamic limit. Here
$N$ is the total number of particles in the system and $V$ is the
volume occupied by the system. Further, we have
\begin{equation}
{\frac {\lambda} {i}} \lim \limits_{{\bf s} \rightarrow 0}
\nabla_s {\cal G} {\left( {\bf x}, {\bf s}; t \right)} = {\frac
{1} {\cal N}} \varrho {\left( {\bf x}; t \right)} {\bf v} {\left(
{\bf x}; t \right)}, \label{eq:velocity}
\end{equation}
and so on.

It is therefore very instructive to derive an equation for the
characteristic function and try to manipulate it in a suitable
manner.

\renewcommand{\theequation}{\thesection.\arabic{equation}}

\setcounter{equation}{0}

\section{Analysis of the Liouville Equation and Mapping onto a
Schr\"{o}dinger Equation}

In this Section, we will work out in detail the simplest case,
where
\begin{equation}
H {\left( {\bf x}, {\bf p}; t \right)} = {\frac {{\bf p}^2} {2 m}}
+ U {\left( {\bf x}; t \right)}. \label{eq:hamiltonian}
\end{equation}
For each term in the Liouville equation (\ref{eq:liouville}), we
have subsequently
\begin{equation}
{\frac {\partial f} {\partial t}} = {\frac {1} {{\left( 2 \pi
\lambda \right)}^n}} \int {\rm d} {\bf s} {\frac {\partial {\cal
G}} {\partial t}} \exp {\left( - {\frac {i} {\lambda}} {\bf s}
\cdot {\bf p} \right)}, \label{eq:firsterm}
\end{equation}
\begin{equation}
{\frac {\bf p} {m}} \cdot \nabla_x f = {\frac {- i \lambda} {m
{\left( 2 \pi \lambda \right)}^n}} \int {\rm d} {\bf s} \nabla_s
\cdot \nabla_x {\cal G} \exp {\left( - {\frac {i} {\lambda}} {\bf
s} \cdot {\bf p} \right)}, \label{eq:secondterm}
\end{equation}
\begin{equation}
\nabla_x U \cdot \nabla_p f = {\frac {- i} {\lambda {\left( 2 \pi
\lambda \right)}^n}} \int {\rm d} {\bf s} {\left( {\bf s} \cdot
\nabla_x U \right)} {\cal G} \exp {\left( - {\frac {i} {\lambda}}
{\bf s} \cdot {\bf p} \right)}. \label{eq:thirdterm}
\end{equation}
Combining all the terms, the sought-for equation can be written
the in form
\begin{equation}
{\frac {\partial {\cal G}} {\partial t}} - {\frac {i \lambda} {m}}
\nabla_s \cdot \nabla_x {\cal G} + {\frac {i} {\lambda}} {\left(
{\bf s} \cdot \nabla_x U \right)} {\cal G} = 0.
\label{eq:eqcharfun}
\end{equation}
We would like now to diagonalize the differential operator,
encountered in the second term of equation (\ref{eq:eqcharfun}).
This is achieved by a simple linear change of variables
\begin{equation}
{\bf x}_1 = {\bf x} + {\frac {{\bf s}} {2}}, \qquad {\bf x}_2 =
{\bf x} - {\frac {{\bf s}} {2}}. \label{eq:changevar}
\end{equation}
Taking into account the identities
\begin{equation}
\nabla_x = \nabla_1 + \nabla_2, \qquad \nabla_s = {\frac {1} {2}}
{\left( \nabla_1 - \nabla_2 \right)}, \label{eq:identities}
\end{equation}
where $\nabla_{1,2}$ denotes the differential operator taken with
respect to the variables ${\bf x}_{1,2}$ respectively, we note
that equation (\ref{eq:eqcharfun}) can be rewritten as
\begin{equation}
{\frac {\partial {\cal G}} {\partial t}} - {\frac {i \lambda} {2
m}} {\left( \nabla_1^2 - \nabla_2^2 \right)} {\cal G} + {\frac {i}
{\lambda}} {\left( {\bf s} \cdot \nabla_x U \right)} {\cal G} = 0.
\label{eq:eqcharfunc}
\end{equation}
The last equation suggests the ansatz
\begin{equation}
{\cal G} {\left( {\bf x}_1, {\bf x}_2; t \right)} = \Psi_1 {\left(
{\bf x}_1; t \right)} \Psi_2 {\left( {\bf x}_2; t \right)}.
\label{eq:ansatz}
\end{equation}
In addition, the yet unknown complex valued functions $\Psi_{k}$
($k = 1, 2$) can be represented in the form
\begin{equation}
\Psi_k {\left( {\bf x}_k; t \right)} = {\cal R}_k {\left( {\bf
x}_k; t \right)} \exp {\left[ {\frac {i} {\lambda}} {\cal S}_k
{\left( {\bf x}_k; t \right)} \right]}, \quad k = 1, 2.
\label{eq:represent}
\end{equation}
Since ${\cal G} {\left( {\bf x}, {\bf s} = 0; t \right)}$ must be
real, it follows immediately that
\begin{equation}
{\cal S} {\left( {\bf x}; t \right)} = {\cal S}_1 {\left( {\bf x};
t \right)} = - {\cal S}_2 {\left( {\bf x}; t \right)}.
\label{eq:condition1}
\end{equation}
Therefore, the characteristic function can be written according to
the relation
\begin{equation}
{\cal G} {\left( {\bf x}, {\bf s}; t \right)} = F {\left( {\bf x},
{\bf s}; t \right)} \exp {\left[ {\frac {i} {\lambda}} G {\left(
{\bf x}, {\bf s}; t \right)} \right]}, \label{eq:charfuncti}
\end{equation}
where
\begin{equation}
F {\left( {\bf x}, {\bf s}; t \right)} = {\cal R}_1 {\left( {\bf
x} + {\frac {\bf s} {2}}; t \right)} {\cal R}_2 {\left( {\bf x} -
{\frac {\bf s} {2}}; t \right)}, \label{eq:amplitude}
\end{equation}
\begin{equation}
G {\left( {\bf x}, {\bf s}; t \right)} = {\cal S} {\left( {\bf x}
+ {\frac {\bf s} {2}}; t \right)} - {\cal S} {\left( {\bf x} -
{\frac {\bf s} {2}}; t \right)}. \label{eq:phase}
\end{equation}
From equation (\ref{eq:phase}) it becomes clear that the phase $G
{\left( {\bf x}, {\bf s}; t \right)}$ is an odd function of the
variable ${\bf s}$. Let us now substitute the characteristic
function represented by (\ref{eq:charfuncti}) into equation
(\ref{eq:eqcharfun}) and separate the real and imaginary part. We
obtain
\begin{equation}
m {\frac {\partial F} {\partial t}} + \nabla_x \cdot {\left( F
\nabla_s G \right)} + {\left( \nabla_s F \right)} \cdot {\left(
\nabla_x G \right)} = 0, \label{eq:realchar}
\end{equation}
\begin{equation}
F {\frac {\partial G} {\partial t}} + {\frac {F} {m}} {\left(
\nabla_s G \right)} \cdot {\left( \nabla_x G \right)} = - F {\bf
s} \cdot \nabla_x U + {\frac {\lambda^2} {m}} \nabla_s \cdot
\nabla_x F. \label{eq:imagchar}
\end{equation}

In what follows, we will analyze equations (\ref{eq:realchar}) and
(\ref{eq:imagchar}) in zero and first order with respect to the
${\bf s}$-variable. The zero-order reads as
\begin{equation}
{\frac {\partial F^{(0)}} {\partial t}} + {\frac {1} {m}} \nabla_x
\cdot {\left( F^{(0)} \nabla_x {\cal S} \right)} = 0,
\label{eq:zeroreal}
\end{equation}
\begin{equation}
{\cal R}_2 \nabla_x {\cal R}_1 - {\cal R}_1 \nabla_x {\cal R}_2 =
0, \label{eq:zeroimag}
\end{equation}
where $F^{(0)} {\left( {\bf x}; t \right)} = {\cal R}_1 {\left(
{\bf x}; t \right)} {\cal R}_2 {\left( {\bf x}; t \right)}$.
Equation (\ref{eq:zeroimag}) simply implies that ${\cal R}_1$ and
${\cal R}_2$ must be equal up to a multiplicative constant that
can be normalized. Thus, without loss of generality, we can write
\begin{equation}
{\cal R} {\left( {\bf x}; t \right)} = {\cal R}_1 {\left( {\bf x};
t \right)} = {\cal R}_2 {\left( {\bf x}; t \right)}.
\label{eq:condition2}
\end{equation}
Therefore, the amplitude $F {\left( {\bf x}, {\bf s}; t \right)}$
is an even function of the ${\bf s}$-variable
\begin{equation}
F {\left( {\bf x}, {\bf s}; t \right)} = {\cal R} {\left( {\bf x}
+ {\frac {\bf s} {2}}; t \right)} {\cal R} {\left( {\bf x} -
{\frac {\bf s} {2}}; t \right)}, \label{eq:amplieven}
\end{equation}
In first order, we obtain
\begin{equation}
\nabla_x {\left( {\frac {\partial {\cal S}} {\partial t}} \right)}
+ \nabla_x {\frac {{\left( \nabla_x {\cal S} \right)}^2} {2m}} = -
\nabla_x U + {\frac {\lambda^2} {2 m}} \nabla_x {\left( {\frac
{\nabla_x^2 {\cal R}} {\cal R}} \right)}, \label{eq:firsimag}
\end{equation}
which integrated once yields
\begin{equation}
{\frac {\partial {\cal S}} {\partial t}} + {\frac {{\left(
\nabla_x {\cal S} \right)}^2} {2m}} = - U + {\frac {\lambda^2} {2
m}} {\frac {\nabla_x^2 {\cal R}} {\cal R}}. \label{eq:firstimag}
\end{equation}
The final step is to introduce the complex wave function $\psi
{\left( {\bf x}; t \right)}$ according to the de Broglie ansatz
\begin{equation}
\psi {\left( {\bf x}; t \right)} = {\cal R} {\left( {\bf x}; t
\right)} \exp {\left[ {\frac {i} {\lambda}} {\cal S} {\left( {\bf
x}; t \right)} \right]}. \label{eq:debroglie}
\end{equation}
Combination of equations (\ref{eq:zeroreal}) and
(\ref{eq:firstimag}) yields the following result
\begin{equation}
i \lambda {\frac {\partial \psi} {\partial t}} = - {\frac
{\lambda^2} {2m}} \nabla_x^2 \psi + U \psi. \label{eq:schrodinger}
\end{equation}
By identification of $\lambda$ with the Planck's constant $\hbar$,
equation (\ref{eq:schrodinger}) transforms into the
Schr\"{o}dinger equation. In addition, equations
(\ref{eq:zeroreal}) and (\ref{eq:firstimag}) coincide with the
system of equations, describing the properties of the Madelung
fluid \cite{Madelung}.

The above considerations can be repeated for the case of
nonrelativistic motion of a spinless particle in electromagnetic
field, governed by the Hamiltonian
\begin{equation}
H {\left( {\bf x}, {\bf p}; t \right)} = {\frac {1} {2 m}} {\left[
{\bf p} - e {\bf A} {\left( {\bf x}; t \right)} \right]}^2 + e U
{\left( {\bf x}; t \right)}. \label{eq:hamiltelmag}
\end{equation}
It is worthwhile to mention that the approach based on the
characteristic function yields the Schr\"{o}dinger equation up to
first order in the variable ${\bf s}$. In this sense it can be
considered as an infinitesimal mapping of the classical kinetic
balance equations onto the Schr\"{o}dinger equation as pointed out
previously \cite{Olavo,Pesci}. However, a drawback of this method
is evident since ansatz (\ref{eq:ansatz}) restricts a possible
class of states, while some classical distributions are not
described with it.

\renewcommand{\theequation}{\thesection.\arabic{equation}}

\setcounter{equation}{0}

\section{Irrotational Momentum Fluctuations}

Our basic assumption concerns the equitability of position and
momentum, which are obviously not on the same footing. We assume
position to be a fundamental variable, while momentum being
proportional to the infinitesimal variation of position respective
to an infinitesimal variation of time cannot be determined
exactly. The physical argument for such assumption is the
following. If an object is perfectly localized, there is no reason
for the impossibility to determine its velocity accurately. If
however, a probability assignment in configuration space strongly
violating particle localization is at hand, there must be some
uncertainty in the specification of the infinitesimal variation of
the particle "position" in the course of time.

Following \cite{Tzenov} instead of (\ref{eq:hamiltonian}), we
consider a dynamical system described by the Hamiltonian
\begin{equation}
H {\left( {\bf x}, {\bf p}; t \right)} = {\frac {1} {2m}} {\left[
{\bf p} + {\bf A} {\left( {\bf x}; t \right)} \right]}^2 + U
{\left( {\bf x}; t \right)}, \label{eq:hamiltonia}
\end{equation}
where ${\bf A} {\left( {\bf x}; t \right)}$ is yet unspecified
fluctuating part of the classical momentum with vanishing mean
value
\begin{equation}
{\left \langle {\bf A} {\left( {\bf x}; t \right)} \right \rangle}
= 0. \label{eq:meanvalue}
\end{equation}
Defining the new variable
\begin{equation}
{\bf P} = {\bf p} + {\bf A}, \label{eq:newvar}
\end{equation}
we can write the Hamilton equations of motion as follows
\begin{equation}
{\frac {{\rm d} {\bf x}} {{\rm d} t}} = {\frac {\bf P} {m}},
\qquad {\frac {{\rm d} {\bf P}} {{\rm d} t}} = {\frac {\partial
{\bf A}} {\partial t}} - \nabla_x U  - {\frac {{\bf P}} {m}}
\times \nabla_x \times {\bf A}. \label{eq:hamilequat}
\end{equation}
Suppose now that ${\bf A}$ is irrotational, that is
\begin{equation}
\nabla_x \times {\bf A} = 0, \label{eq:curlfree}
\end{equation}
which also implies
\begin{equation}
{\bf A} = - \nabla_x \Phi. \label{eq:curlfreee}
\end{equation}
With these observations, it follows that the Hamilton equations
(\ref{eq:hamilequat}) can be obtained from a new Hamiltonian
\begin{equation}
{\cal H} {\left( {\bf x}, {\bf P}; t \right)} = {\frac {{\bf P}^2}
{2m}} + U {\left( {\bf x}; t \right)} + {\frac {\partial \Phi
{\left( {\bf x}; t \right)}} {\partial t}}, \label{eq:newhamil}
\end{equation}

Next, we perform a polynomial marginalization if the distribution
function $f {\left( {\bf x}, {\bf P}; t \right)}$. This is done by
initially multiplying the stochastic Liouville equation
\begin{equation}
{\frac {\partial f} {\partial t}} + {\frac {\bf P} {m}} \cdot
\nabla_x f + {\bf F} \cdot \nabla_{\bf P} f = 0
\label{eq:liouvequat}
\end{equation}
by various powers $P_1^{k_1} P_2^{k_2} \dots P_n^{k_n}$ and then
formally integrating over the momentum variable. Here, the random
force ${\bf F}$ is given by the expression
\begin{equation}
{\bf F} = {\frac {\partial {\bf A}} {\partial t}} - \nabla_x U.
\label{eq:force}
\end{equation}
The equations for the first two moments can be written as
\begin{equation}
{\frac {\partial \varrho} {\partial t}} + \nabla_x \cdot {\left(
\varrho {\bf V} \right)} = 0, \label{eq:continuity}
\end{equation}
\begin{equation}
{\frac {\partial {\left( \varrho V_n \right)}} {\partial t}} +
{\frac {\partial \Pi_{kn}} {\partial x_k}} = {\frac {\varrho F_n}
{m}}, \label{eq:momentum}
\end{equation}
where
\begin{equation}
\varrho {\left( {\bf x}; t \right)} = m {\cal N} \int {\rm d} {\bf
P} f {\left( {\bf x}, {\bf P}; t \right)}, \label{eq:densit}
\end{equation}
\begin{equation}
\varrho {\left( {\bf x}; t \right)} {\bf V} {\left( {\bf x}; t
\right)} = {\cal N} \int {\rm d} {\bf P} {\bf P} f {\left( {\bf
x}, {\bf P}; t \right)}, \label{eq:currentvel}
\end{equation}
\begin{equation}
\Pi_{kl} {\left( {\bf x}; t \right)} = {\frac {\cal N} {m}} \int
{\rm d} {\bf P} P_k P_l f {\left( {\bf x}, {\bf P}; t \right)}.
\label{eq:stress}
\end{equation}
It is a simple matter to verify that the  Liouville equation
(\ref{eq:liouvequat}) possesses an exact solution of the form
\begin{equation}
f {\left( {\bf x}, {\bf P}; t \right)} = {\frac {\varrho {\left(
{\bf x}; t \right)}} {m {\cal N}}} \delta {\left[ {\bf P} - m {\bf
V} {\left( {\bf x}; t \right)} \right]}, \label{eq:exacrsolut}
\end{equation}
which is also known as the classical Bohm distribution, usually
interpreted as the classical limit of quantum pure states
\cite{Prata2}. Substitution of the classical Bohm distribution
into equation (\ref{eq:stress}) yields the expression
\begin{equation}
\Pi_{kl} = \varrho V_k V_l, \label{eq:stressten}
\end{equation}
for the stress tensor $\Pi_{kl}$. Hence the system
(\ref{eq:continuity}) and (\ref{eq:momentum}) represents an exact
closure of hydrodynamic equations, fully equivalent to the
Liouville equation.

We now take into account the fact that the momentum ${\bf P}$ and
hence the current velocity ${\bf V}$ consists of a classical mean
part ${\bf v}$ (corresponding to ${\bf p}$) and a fluctuation
${\widetilde{\bf V}}$ (corresponding to ${\bf A}$). Averaging
equations (\ref{eq:continuity}) and (\ref{eq:momentum}), we obtain
\begin{equation}
{\frac {\partial \varrho} {\partial t}} + \nabla_x \cdot {\left(
\varrho {\bf v} \right)} = 0, \label{eq:continuit}
\end{equation}
\begin{equation}
{\frac {\partial v_n} {\partial t}} + v_k {\frac {\partial v_n}
{\partial x_k}} = - {\frac {1} {m}} {\frac {\partial U} {\partial
x_n}} - {\frac {1} {\varrho}} {\frac {\partial} {\partial x_k}}
{\left( \varrho {\cal C}_{kn} \right)}, \label{eq:momentu}
\end{equation}
where
\begin{equation}
{\cal C}_{kn} = {\left \langle {\widetilde{V}}_k {\widetilde{V}}_n
\right \rangle}. \label{eq:tensor}
\end{equation}

Recall that it was initially assumed that the momentum
fluctuations are irrotational. This also implies irrotationality
of the second term on the right-hand-side of equation
(\ref{eq:momentu}). It can be written in the form
\begin{equation}
{\frac {\partial {\cal Z}} {\partial x_n}} = {\frac {\partial
{\cal C}_{kn}} {\partial x_k}} + {\cal C}_{kn} {\frac {\partial R}
{\partial x_k}}, \label{eq:irrotat}
\end{equation}
where
\begin{equation}
R = \ln \varrho. \label{eq:rvariable}
\end{equation}
Taking curl of both sides of equation (\ref{eq:irrotat}), we
obtain
\begin{equation}
\epsilon_{lmn} {\frac {\partial} {\partial x_m}} {\left( {\frac
{\partial {\cal C}_{kn}} {\partial x_k}} + {\cal C}_{kn} {\frac
{\partial R} {\partial x_k}} \right)} = 0, \label{eq:irrotati}
\end{equation}
where as usual, $\epsilon_{lmn}$ denotes the fully antisymmetric
third-rank unit tensor. Multiplication by $\epsilon_{lqp}$ and
summation on $l$ in the last identity, yields a second order
linear partial differential equation for the unknown correlation
tensor ${\cal C}_{kn}$
\begin{widetext}
\begin{equation}
{\left( {\frac {\partial} {\partial x_k}} + {\frac {\partial R}
{\partial x_k}} \right)} {\left( {\frac {\partial {\cal C}_{kn}}
{\partial x_m}} - {\frac {\partial {\cal C}_{km}} {\partial x_n}}
\right)} + {\cal C}_{kn} {\frac {\partial^2 R} {\partial x_k
\partial x_m}} - {\cal C}_{km} {\frac {\partial^2 R}
{\partial x_k \partial x_n}} = 0. \label{eq:irrotatio}
\end{equation}
\end{widetext}
Note that the left-hand-side of equation (\ref{eq:irrotatio}) is
antisymmetric with respect to the indices $m$ and $n$. This
restricts considerably the number of its solutions. On the other
hand, equation (\ref{eq:irrotatio}) is a linear equation with
respect to ${\cal C}_{kn}$, so that its general solution can be
written as a linear combination of particular solutions. Since the
correlation tensor is symmetric, it can be represented in diagonal
form. To find a particular solution, suppose that equation
(\ref{eq:irrotatio}) is written in a reference frame in which the
correlation tensor is diagonal. Then, all elements must be equal,
which is the only possibility for this particular solution
[represented by the second term in equation (\ref{eq:generform})].
Another solution can be represented in the form of a Hessian
matrix of a generic function. A simple and straightforward
verification shows that these two particular solutions exhaust all
possibilities, and ${\cal C}_{kn}$ can be written as follows
\begin{equation}
{\cal C}_{kn} = \alpha {\frac {\partial^2 \Gamma} {\partial x_k
\partial x_n}} + \beta \delta_{kn} {\cal F}, \label{eq:generform}
\end{equation}
where $\alpha$ and $\beta$ are constant coefficients, and the
scalar functions ${\cal F} {\left( {\bf x}; t \right)}$ and
$\Gamma {\left( {\bf x}; t \right)}$ are some functions of ${\bf
x}$ and $t$ that remain to be determined. Direct substitution of
the solution (\ref{eq:generform}) into equation
(\ref{eq:irrotatio}) shows that ${\cal F}$ is an arbitrary
function of $R$
\begin{equation}
{\cal F} {\left( {\bf x}; t \right)} = F {\left( R \right)},
\label{eq:ffunction}
\end{equation}
while $\Gamma$ is equal to $R$ up to a multiplicative constant,
which (without loss of generality) can be set equal to unity
\begin{equation}
\Gamma {\left( {\bf x}; t \right)} = R {\left( {\bf x}; t
\right)}. \label{eq:gfunction}
\end{equation}
Hence, equation (\ref{eq:irrotat}) can be rewritten as follows
\begin{equation}
\nabla_x {\cal Z} = \alpha \nabla_x {\left[ \nabla_x^2 R + {\frac
{1} {2}} {\left( \nabla_x R \right)}^2 \right]} + \nabla_x {\cal
F} + {\cal F} \nabla_x R \nonumber
\end{equation}
\begin{equation}
= 2 \alpha \nabla_x {\left( {\frac {\nabla_x^2 {\sqrt{\varrho}}}
{\sqrt{\varrho}}} \right)} + \nabla_x {\cal F} + {\cal F} \nabla_x
R. \label{eq:finalpotent}
\end{equation}

We would like now to show that the arbitrary constant $\alpha$
must be negative ($\alpha < 0$). In analogy to the definitions
(\ref{eq:densit})--(\ref{eq:stress}), we can introduce the kinetic
energy density according to the relation
\begin{equation}
E {\left( {\bf x}; t \right)} = {\frac {\cal N} {2m}} \int {\rm d}
{\bf P} {\bf P}^2 f {\left( {\bf x}, {\bf P}; t \right)} = {\frac
{\varrho {\bf V}^2} {2}}. \label{eq:energy}
\end{equation}
Averaging equation (\ref{eq:energy}), we obtain
\begin{equation}
{\left \langle E {\left( {\bf x}; t \right)} \right \rangle} =
{\frac {\varrho {\bf v}^2} {2}} + {\frac {\varrho} {2}} {\rm Tr}
{\cal C}, \label{eq:averenergy}
\end{equation}
where ${\rm Tr} {\cal C}$ denotes the trace of the correlation
tensor (\ref{eq:tensor}). The second term on the right-hand-side
of equation (\ref{eq:averenergy}) represents the density of the
internal energy, which is due to the fluctuating part of the
current velocity. The total internal energy is given by the
expression
\begin{equation}
{\cal E} = {\frac {1} {2}} \int {\rm d} {\bf x} \varrho {\rm Tr}
{\cal C} = {\frac {n} {2}} \int {\rm d} {\bf x} \varrho {\cal F}
{\left( R \right)} - {\frac {\alpha} {2}} \int {\rm d} {\bf x}
{\frac {{\left( \nabla_x \varrho \right)}^2} {\varrho}},
\label{eq:intenergy}
\end{equation}
where integration by parts and taking into account of vanishing
integrals has been performed in the second term on the
right-hand-side. Since the total internal energy must be positive
for any choice of the arbitrary function ${\cal F}$ (including
${\cal F} = 0$), the free parameter $\alpha$ must be negative.
Remarkably enough, the second term on the right-hand-side of
equation (\ref{eq:intenergy}) is proportional to the Fisher
information \cite{Kullback}
\begin{equation}
{\cal I} = \int {\rm d} {\bf x} {\frac {{\left( \nabla_x \varrho
\right)}^2} {\varrho}}. \label{eq:fisherinf}
\end{equation}
The above equation (\ref{eq:fisherinf}) represents a direct link
between quantum mechanics and Fisher information theory. It is
remarkable that in the simplest case, where ${\cal F} = 0$, the
Fisher information measure ${\cal I}$ has a very transparent
meaning as being simply a measure of the internal energy of the
quantum system.

First of all, we would like to explore the simplest case, where
${\cal F} = 0$. Identifying the parameter $\alpha$ as
\begin{equation}
\alpha = - {\frac {\hbar^2} {4m^2}}, \label{eq:paramalpha}
\end{equation}
equation (\ref{eq:momentu}) can be written accordingly
\begin{equation}
{\frac {\partial {\bf v}} {\partial t}} + {\bf v} \cdot \nabla_x
{\bf v} = - {\frac {\nabla_x U} {m}} + {\frac {\hbar^2} {2m^2}}
\nabla_x {\left( {\frac {\nabla_x^2 {\sqrt{\varrho}}}
{\sqrt{\varrho}}} \right)}. \label{eq:momentuqm}
\end{equation}
If we further define
\begin{equation}
{\bf v} = {\frac {1} {m}} \nabla_x {\cal S}, \label{eq:eikonal}
\end{equation}
the de Broglie ansatz (\ref{eq:debroglie}) with
\begin{equation}
\lambda = \hbar, \qquad \qquad {\cal R} = {\sqrt{\varrho}},
\label{eq:define}
\end{equation}
yields immediately the Schr\"{o}dinger equation
\begin{equation}
i \hbar {\frac {\partial \psi} {\partial t}} = - {\frac {\hbar^2}
{2m}} \nabla_x^2 \psi + U \psi. \label{eq:schrodinge}
\end{equation}

Another simple but nontrivial example is the case where ${\cal F}
= b / m = {\rm const}$. As a result, we obtain the Schr\"{o}dinger
equation with logarithmic nonlinearity \cite{Birula}
\begin{equation}
i \hbar {\frac {\partial \psi} {\partial t}} = - {\frac {\hbar^2}
{2m}} \nabla_x^2 \psi + {\left( U + b \ln {\left| \psi \right|}^2
\right)} \psi. \label{eq:schrodlog}
\end{equation}
Note that the constant $b$ can be positive, as well as negative.
From the requirement of the positivity of the internal energy, it
is easy to obtain the upper bound for the case $b < 0$
\begin{equation}
{\left| b \right|} < {\frac {\hbar^2} {4n m^2 N}} {\cal I}.
\label{eq:upperbound}
\end{equation}

Concluding this Section, it is worthwhile to reiterate that the
Schr\"{o}dinger equation has been derived by the sole use of
purely classical stochastic arguments. In addition, it should be
pointed out that the linear Schr\"{o}dinger equation is not the
unique possibility.

\renewcommand{\theequation}{\thesection.\arabic{equation}}

\setcounter{equation}{0}

\section{The Quantum Picture}

First of all, we note that $\varrho$, which has the meaning of
mass density [see equation (\ref{eq:densit})] can be rescaled
$\varrho \rightarrow mN \varrho$, such that it becomes normalized
\begin{equation}
\int {\rm d} {\bf x} \varrho {\left( {\bf x}; t \right)} = 1.
\label{eq:normalize}
\end{equation}
This implies that the wave function is normalized as well
\begin{equation}
\int {\rm d} {\bf x} {\left| \psi {\left( {\bf x}; t \right)}
\right|}^2 = \int {\rm d} {\bf x} \varrho {\left( {\bf x}; t
\right)} = 1. \label{eq:normalwavef}
\end{equation}
From (\ref{eq:debroglie}) and (\ref{eq:exacrsolut}) for the
expectation value of an arbitrary function $F {\left( {\bf x}
\right)}$ of position, we obtain
\begin{equation}
{\left \langle F {\left( {\bf x} \right)} \right \rangle} = \int
{\rm d} {\bf x} {\rm d} {\bf P} F {\left( {\bf x} \right)} f
{\left( {\bf x}, {\bf P}; t \right)} = \int {\rm d} {\bf x} F
{\left( {\bf x} \right)} {\left| \psi {\left( {\bf x}; t \right)}
\right|}^2. \label{eq:expeposit}
\end{equation}
The expectation value of momentum can be found in a similar manner
\begin{equation}
{\left \langle {\bf P} \right \rangle} = \int {\rm d} {\bf x} {\rm
d} {\bf P} {\bf P} f {\left( {\bf x}, {\bf P}; t \right)} = m \int
{\rm d} {\bf x} {\left \langle {\bf V} \right \rangle} \varrho
{\left( {\bf x}; t \right)} \nonumber
\end{equation}
\begin{equation}
= - i \hbar \int {\rm d} {\bf x} \psi^{\ast} \nabla_x \psi.
\label{eq:expemoment}
\end{equation}
Further, we have (for the case, where ${\cal F} = 0$)
\begin{equation}
{\left \langle {\bf P}^2 \right \rangle} = m^2 \int {\rm d} {\bf
x} \varrho {\left( {\bf x}; t \right)} {\left( {\bf v}^2 + {\rm
Tr} {\cal C} \right)} \nonumber
\end{equation}
\begin{equation}
= \hbar^2 \int {\rm d} {\bf x} {\left( \nabla_x \psi^{\ast}
\right)} \cdot {\left( \nabla_x \psi \right)}.
\label{eq:expemoments}
\end{equation}
The expectation value for the energy is represented by the
expression (for the case, where ${\cal F} = 0$)
\begin{equation}
{\left \langle {\cal H} \right \rangle} = \int {\rm d} {\bf x}
{\left[ {\frac {\hbar^2} {2m}} {\left( \nabla_x \psi^{\ast}
\right)} \cdot {\left( \nabla_x \psi \right)} + U {\left( {\bf x};
t \right)} {\left| \psi \right|}^2 \right]}. \label{eq:expecener}
\end{equation}
Thus, we recover the basic quantum rules to calculate expectation
values of observables, which are not higher than quadratic in
momentum. This implies that not only expectation values, but also
uncertainties for position and momentum can be calculated so that
they correspond to the standard quantum expression. Similar to the
observations made by M.J.W. Hall and M. Reginatto
\cite{Hall,Reginatto}, we find that the Heisenberg uncertainty
principle can be obtained solely by using classical statistical
mechanics formalism with stochastic ingredients added.

Since only the mean value of the current velocity ${\bf v}$ and
the correlation tensor ${\cal C}_{mn}$ have been specified to this
end, it is not immediately clear how one can proceed with
calculation of expectation values of an arbitrary function of
momentum. A possible approach to this problem will be outlined in
the sequel.

The definition (\ref{eq:charfunc}) of the characteristic function
can be generalized into a characteristic functional
\begin{equation}
{\left \langle {\cal G} {\left( {\bf x}, {\bf s}; t \right)}
\right \rangle} = {\left| \psi {\left( {\bf x}; t \right)}
\right|}^2 {\left \langle \exp {\left( {\frac {im} {\hbar}} {\bf
s} \cdot {\bf V} \right)} \right \rangle}. \label{eq:functional}
\end{equation}
On the other hand, from (\ref{eq:ansatz}) we would obtain
\begin{equation}
{\left \langle {\cal G} {\left( {\bf x}, {\bf s}; t \right)}
\right \rangle} = \psi {\left( {\bf x} + {\frac {\bf s} {2}}; t
\right)} \psi^{\ast} {\left( {\bf x} - {\frac {\bf s} {2}}; t
\right)}, \label{eq:functwigner}
\end{equation}
which implies
\begin{equation}
{\left| \psi {\left( {\bf x}; t \right)} \right|}^2 {\left \langle
\exp {\left( {\frac {im} {\hbar}} {\bf s} \cdot {\bf V} \right)}
\right \rangle} = \psi {\left( {\bf x} + {\frac {\bf s} {2}}; t
\right)} \psi^{\ast} {\left( {\bf x} - {\frac {\bf s} {2}}; t
\right)}. \label{eq:functionalw}
\end{equation}
The last equation represents a strong condition to be imposed on
the stochastic properties of the random velocity field. Clearly,
higher order correlation functions of the random velocity field
${\widetilde{\bf V}}$ must be specified accordingly in order to
satisfy equation (\ref{eq:functionalw}). The consequences are
analyzed in the Appendix. There, it is shown that equation
(\ref{eq:functionalw}) remains valid to second order in the
variable ${\bf s}$. In addition, an expression for the third-order
correlation function ${\left \langle {\widetilde{V}}_k
{\widetilde{V}}_l {\widetilde{V}}_n \right \rangle}$ of the random
velocity field ${\widetilde{\bf V}}$ is derived.

Let us define the Fourier transform of the wave function according
to the well-known relation
\begin{equation}
\psi {\left( {\bf x}; t \right)} = {\frac {1} {{\left( 2 \pi \hbar
\right)}^{n/2}}} \int {\rm d} {\bf p} {\widehat{\psi}} {\left(
{\bf p}; t \right)} \exp {\left( {\frac {i} {\hbar}} {\bf x} \cdot
{\bf p} \right)}. \label{eq:fouriertrans}
\end{equation}
In order to cast a parallel with the discussion in Section II, we
integrate equation (\ref{eq:functwigner}) over ${\bf x}$. As a
result, we obtain
\begin{equation}
{\overline{\cal G}} (s) = \int {\rm d} {\bf x} {\left \langle
{\cal G} {\left( {\bf x}, {\bf s}; t \right)} \right \rangle} =
\int {\rm d} {\bf p} {\left| {\widehat{\psi}} {\left( {\bf p}; t
\right)} \right|}^2 \exp {\left( {\frac {i} {\hbar}} {\bf s} \cdot
{\bf p} \right)}. \label{eq:genfexpval}
\end{equation}
According to expressions (\ref{eq:density}) and
(\ref{eq:velocity}) and their obvious generalization, the
expectation value of an arbitrary function of momentum $F {\left(
{\bf p} \right)}$ can be written in the form
\begin{equation}
{\left \langle F {\left( {\bf p} \right)} \right \rangle} =
{\left. {\widehat{F}} {\left( -i \hbar \nabla_s \right)}
{\overline{\cal G}} (s) \right|}_{{\bf s}=0} = \int {\rm d} {\bf
p} F {\left( {\bf p} \right)} {\left| {\widehat{\psi}} {\left(
{\bf p}; t \right)} \right|}^2. \label{eq:expvalmom}
\end{equation}
In coordinate representation, we have
\begin{equation}
{\left \langle F {\left( {\bf p} \right)} \right \rangle} = \int
{\rm d} {\bf x} \psi^{\ast} {\left( {\bf x}; t \right)}
{\widehat{F}} {\left( -i \hbar \nabla_x \right)} \psi {\left( {\bf
x}; t \right)}. \label{eq:expvalmomeqf}
\end{equation}
A natural generalization is now in order
\begin{equation}
{\left \langle F {\left( {\bf x}, {\bf p} \right)} \right \rangle}
= \int {\rm d} {\bf x} \psi^{\ast} {\left( {\bf x}; t \right)}
{\widehat{F}} {\left( {\bf x}, -i \hbar \nabla_x \right)} \psi
{\left( {\bf x}; t \right)}, \label{eq:expvalcoormom}
\end{equation}
where as usual, an appropriate operator ordering must be
specified.

We end this Section by emphasizing another remarkable link between
the formalism developed here and the Wigner-Weyl approach. Using
the compatibility condition in the form of (\ref{eq:functwigner}),
we can rewrite equation (\ref{eq:charfunct}) as
\begin{widetext}
\begin{equation}
{\left \langle f {\left( {\bf x}, {\bf p}; t \right)} \right
\rangle} = {\frac {1} {{\left( 2 \pi \hbar \right)}^n}} \int {\rm
d} {\bf s} \psi {\left( {\bf x} + {\frac {\bf s} {2}}; t \right)}
\psi^{\ast} {\left( {\bf x} - {\frac {\bf s} {2}}; t \right)} \exp
{\left( - {\frac {i} {\hbar}} {\bf s} \cdot {\bf p} \right)},
\label{eq:charfwigner}
\end{equation}
\end{widetext}
This implies that ${\left| \psi \right|}^2 {\left \langle \delta
{\left( {\bf p} - m {\bf V} \right)} \right \rangle}$ is the
Wigner function. Leaving more speculations aside on the fact that
the latter is a quasi-distribution which is not always positive,
we note that equation (\ref{eq:charfwigner}) represents a relation
between the averaged classical Bohm distribution and the Wigner
function.

\renewcommand{\theequation}{\thesection.\arabic{equation}}

\setcounter{equation}{0}

\section{Moyal Bracket}

The Moyal bracket is a useful tool when one wishes to determine a
semiclassical limit to wave mechanics. Moyal \cite{Moyal}
elaborated on the theory of Wigner \cite{Wigner} on how to
describe quantum systems in phase space in a way which is formally
analogous to the dynamics of classical distributions. The Moyal
bracket provides a semiclassical limit to quantum mechanical
commutation relations, which is what is of interest to us, and we
consider this in some detail in the present Section.

First of all, let us introduce the characteristic dynamical
variable defined as
\begin{equation}
{\cal C} {\left( {\bf x}, {\bf p}; {\bf k}, {\bf s} \right)} =
\exp {\left[ {\frac {i} {\hbar}} {\left( {\bf k} \cdot {\bf x} +
{\bf s} \cdot {\bf p} \right)} \right]}. \label{eq:chardynvaroper}
\end{equation}
The rule to calculate expectation values according to expression
(\ref{eq:expvalcoormom}) suggests the introduction of the
corresponding characteristic operator \cite{Gardiner}
\begin{equation}
{\widehat{\cal C}} {\left( {\widehat{\bf x}}, {\widehat{\bf p}};
{\bf k}, {\bf s} \right)} = \exp {\left[ {\frac {i} {\hbar}}
{\left( {\bf k} \cdot {\widehat{\bf x}} + {\bf s} \cdot
{\widehat{\bf p}} \right)} \right]}, \label{eq:charoper}
\end{equation}
where ${\widehat{\bf x}}$ implies ${\bf x}$, while ${\widehat{\bf
p}} = -i \hbar \nabla_x$. The characteristic function is defined
as the expectation value
\begin{equation}
{\widetilde{\cal C}} {\left( {\bf k}, {\bf s} \right)} = {\left
\langle {\cal C} {\left( {\bf x}, {\bf p}; {\bf k}, {\bf s}
\right)} \right \rangle}. \label{eq:charfunctoper}
\end{equation}
According to equation (\ref{eq:expvalcoormom}) and taking into
account the Campbell-Baker-Hausdorff identity,
\begin{equation}
e^{{\widehat{A}} + {\widehat{B}}} = e^{{\widehat{A}}}
e^{{\widehat{B}}} e^{- {\frac {1} {2}} {\left[ {\widehat{A}},
{\widehat{B}} \right]}}, \label{eq:cbhidentity}
\end{equation}
when the commutator commutes with both ${\widehat{A}}$ and
${\widehat{B}}$ (which is the case for operators proportional to
${\widehat{\bf x}}$ and ${\widehat{\nabla}}_x$), we obtain
\begin{equation}
{\widetilde{\cal C}} {\left( {\bf k}, {\bf s} \right)} = \int {\rm
d} {\bf x} \psi {\left( {\bf x} + {\frac {\bf s} {2}}; t \right)}
\psi^{\ast} {\left( {\bf x} - {\frac {\bf s} {2}}; t \right)} \exp
{\left( {\frac {i} {\hbar}} {\bf k} \cdot {\bf x} \right)}.
\label{eq:charfunctop}
\end{equation}
Taking into account the inverse Fourier transform of expression
(\ref{eq:charfunctop}) and equation (\ref{eq:charfwigner}), it
immediately follows that the characteristic function is a double
Fourier transform of the Wigner function
\begin{widetext}
\begin{equation}
W {\left( {\bf x}, {\bf p}; t \right)} = {\left \langle f {\left(
{\bf x}, {\bf p}; t \right)} \right \rangle} = {\frac {1} {{\left(
2 \pi \hbar \right)}^{2n}}} \int {\rm d} {\bf k} {\rm d} {\bf s}
{\widetilde{\cal C}} {\left( {\bf k}, {\bf s} \right)} \exp
{\left[ - {\frac {i} {\hbar}} {\left( {\bf k} \cdot {\bf x} + {\bf
s} \cdot {\bf p} \right)} \right]}, \label{eq:charfunwigner}
\end{equation}
\end{widetext}
or
\begin{equation}
{\widetilde{\cal C}} {\left( {\bf k}, {\bf s} \right)} = \int {\rm
d} {\bf x} {\rm d} {\bf p} W {\left( {\bf x}, {\bf p}; t \right)}
\exp {\left[ {\frac {i} {\hbar}} {\left( {\bf k} \cdot {\bf x} +
{\bf s} \cdot {\bf p} \right)} \right]}. \label{eq:charfunwigner1}
\end{equation}
Equation (\ref{eq:charfunwigner1}) implies also that the
expectation value of the characteristic dynamical variable
(\ref{eq:chardynvaroper}) is a result of integration of its
product with the Wigner function over all of phase space.

Further, a generic dynamical variable ${A} {\left( {\bf x}, {\bf
p} \right)}$ is represented by the expression
\begin{equation}
{A} {\left( {\bf x}, {\bf p} \right)} = {\frac {1} {{\left( 2 \pi
\hbar \right)}^{2n}}} \int {\rm d} {\bf k} {\rm d} {\bf s} {\cal
A} {\left( {\bf k}, {\bf s} \right)} {\cal C} {\left( {\bf x},
{\bf p}; {\bf k}, {\bf s} \right)}, \label{eq:gendynvar}
\end{equation}
while its corresponding operator in terms of ${\widehat{\bf x}}$
and ${\widehat{\bf p}}$ specified by the characteristic operator
(\ref{eq:charoper}) can be written as
\begin{equation}
{\widehat{A}} {\left( {\widehat{\bf x}}, {\widehat{\bf p}}
\right)} = {\frac {1} {{\left( 2 \pi \hbar \right)}^{2n}}} \int
{\rm d} {\bf k} {\rm d} {\bf s} {\cal A} {\left( {\bf k}, {\bf s}
\right)} {\widehat{\cal C}} {\left( {\widehat{\bf x}},
{\widehat{\bf p}}; {\bf k}, {\bf s} \right)}. \label{eq:genopevar}
\end{equation}
The dynamical variable (\ref{eq:gendynvar}) is usually called a
phase function of the operator (\ref{eq:genopevar}). We can take
the Fourier transform of the phase function $A {\left( {\bf x};
{\bf p} \right)}$, which returns ${\cal A} {\left( {\bf k}; {\bf
s} \right)}$, and substitute it back into equation
(\ref{eq:genopevar}) for the corresponding operator. This results
in
\begin{widetext}
\begin{equation}
{\widehat{A}} {\left( {\widehat{\bf x}}, {\widehat{\bf p}}
\right)} = {\frac {1} {{\left( 2 \pi \hbar \right)}^{2n}}} \int
{\rm d} {\bf x} {\rm d} {\bf p} {\rm d} {\bf k} {\rm d} {\bf s} A
{\left( {\bf x}; {\bf p} \right)} \exp {\left\{ - {\frac {i}
{\hbar}} {\left[ {\bf k} \cdot {\left( {\bf x} - {\widehat{\bf x}}
\right)} + {\bf s} \cdot {\left( {\bf p} - {\widehat{\bf p}}
\right)} \right]} \right\}}. \label{eq:genopevar1}
\end{equation}
\end{widetext}
Taking into account expression (\ref{eq:charfunwigner1}) for the
expectation value of the characteristic dynamical variable, we
obtain
\begin{equation}
{\left \langle A {\left( {\bf x}, {\bf p} \right)} \right \rangle}
= {\left \langle {\widehat{A}} {\left( {\widehat{\bf x}},
{\widehat{\bf p}} \right)} \right \rangle} = \int {\rm d} {\bf x}
{\rm d} {\bf p} A {\left( {\bf x}, {\bf p} \right)} W {\left( {\bf
x}, {\bf p}; t \right)}. \label{eq:expectvalue}
\end{equation}
This describes the expectation value of a dynamical variable (an
operator observable in quantum mechanical sense) as being the
result of integrating its product (the product of its
corresponding phase function) with the Wigner function over all of
phase space. In this way the Wigner function acts much like a
joint probability distribution over position and momentum.

Expression (\ref{eq:expvalcoormom}) shows that the expectation
value of the product of two dynamical variables [corresponding
operators in the sense of equation (\ref{eq:genopevar})]depends on
the order of the multipliers, and in this sense they do not
commute. We now wish to determine the measure of non commutativity
of two arbitrary operators
\begin{equation}
i \hbar {\widehat{D}} {\left( {\widehat{\bf x}}, {\widehat{\bf p}}
\right)} = {\left[ {\widehat{A}} {\left( {\widehat{\bf x}},
{\widehat{\bf p}} \right)}, {\widehat{B}} {\left( {\widehat{\bf
x}}, {\widehat{\bf p}} \right)} \right]}, \label{eq:commutator}
\end{equation}
which is expected to be proportional to $\hbar$. We begin by
substituting in the general expression of equation
(\ref{eq:genopevar}) for each of the operators ${\widehat{A}}
{\left( {\widehat{\bf x}}, {\widehat{\bf p}} \right)}$,
${\widehat{B}} {\left( {\widehat{\bf x}}, {\widehat{\bf p}}
\right)}$ and ${\widehat{D}} {\left( {\widehat{\bf x}},
{\widehat{\bf p}} \right)}$. This brings us to
\begin{widetext}
\begin{equation}
i \hbar \int {\rm d} {\bf k} {\rm d} {\bf s} {\cal D} {\left( {\bf
k}, {\bf s} \right)} {\widehat{\cal C}} {\left( {\bf k}, {\bf s}
\right)} = {\frac {1} {{\left( 2 \pi \hbar \right)}^{2n}}} \int
{\rm d} {\bf k}_1 {\rm d} {\bf s}_1 {\rm d} {\bf k}_2 {\rm d} {\bf
s}_2 {\cal A} {\left( {\bf k}_1, {\bf s}_1 \right)} {\cal B}
{\left( {\bf k}_2, {\bf s}_2 \right)} {\left[ {\widehat{\cal C}}
{\left( {\bf k}_1, {\bf s}_1 \right)}, {\widehat{\cal C}} {\left(
{\bf k}_2, {\bf s}_2 \right)} \right]}, \label{eq:identity}
\end{equation}
\end{widetext}
where for the sake of simplicity the explicit dependence on the
operators ${\widehat{\bf x}}$ and ${\widehat{\bf p}}$ has been
dropped. Taking into account the Campbell-Baker-Hausdorff identity
(\ref{eq:cbhidentity}), and the fact that the commutator
\begin{equation}
{\left[ {\bf k}_1 \cdot {\widehat{\bf x}} + {\bf s}_1 \cdot
{\widehat{\bf p}}, {\bf k}_2 \cdot {\widehat{\bf x}} + {\bf s}_2
\cdot {\widehat{\bf p}} \right]} = - i \hbar {\left( {\bf k}_2
\cdot {\bf s}_1 - {\bf k}_1 \cdot {\bf s}_2 \right)},
\label{eq:commutator1}
\end{equation}
is a scalar quantity it is straightforward to verify
\begin{equation}
{\left[ {\widehat{\cal C}} {\left( {\bf k}_1, {\bf s}_1 \right)},
{\widehat{\cal C}} {\left( {\bf k}_2, {\bf s}_2 \right)} \right]}
\nonumber
\end{equation}
\begin{equation}
= 2i {\widehat{\cal C}} {\left( {\bf k}_1 + {\bf k}_2, {\bf s}_1 +
{\bf s}_2 \right)} \sin {\frac {{\bf k}_2 \cdot {\bf s}_1 - {\bf
k}_1 \cdot {\bf s}_2} {2 \hbar}}. \label{eq:commutator2}
\end{equation}
We now determine the expectation value on both sides of equation
(\ref{eq:identity}), resulting in characteristic functions on each
side of the resulting equation. Substituting expression
(\ref{eq:charfunwigner1}) for the characteristic function and
using the representation (\ref{eq:gendynvar}), we obtain
\begin{equation}
\hbar \int {\rm d} {\bf x} {\rm d} {\bf p} D {\left( {\bf x}, {\bf
p} \right)} W {\left( {\bf x}, {\bf p}; t \right)} \nonumber
\end{equation}
\begin{widetext}
\begin{equation}
= {\frac {2} {{\left( 2 \pi \hbar \right)}^{4n}}} \int {\rm d}
{\bf x} {\rm d} {\bf p} {\rm d} {\bf k}_1 {\rm d} {\bf s}_1 {\rm
d} {\bf k}_2 {\rm d} {\bf s}_2 W {\left( {\bf x}, {\bf p}; t
\right)} {\cal A} {\left( {\bf k}_1, {\bf s}_1 \right)} {\cal B}
{\left( {\bf k}_2, {\bf s}_2 \right)} \sin {\frac {{\bf k}_2 \cdot
{\bf s}_1 - {\bf k}_1 \cdot {\bf s}_2} {2 \hbar}} {\cal C} {\left(
{\bf k}_1, {\bf s}_1 \right)} {\cal C} {\left( {\bf k}_2, {\bf
s}_2 \right)}. \label{eq:identity2}
\end{equation}
\end{widetext}
Next, we note that the argument of the sine on the right-hand-side
of equation (\ref{eq:identity2}) can be represented as
\begin{equation}
{\left( {\bf k}_2 \cdot {\bf s}_1 - {\bf k}_1 \cdot {\bf s}_2
\right)} {\cal C} {\left( {\bf k}_1, {\bf s}_1 \right)} {\cal C}
{\left( {\bf k}_2, {\bf s}_2 \right)} \nonumber
\end{equation}
\begin{equation}
= \hbar^2 {\left( \nabla_{x1} \cdot \nabla_{p2} - \nabla_{p1}
\cdot \nabla_{x2} \right)} {\cal C} {\left( {\bf k}_1, {\bf s}_1
\right)} {\cal C} {\left( {\bf k}_2, {\bf s}_2 \right)},
\label{eq:represine}
\end{equation}
where the subscripts on the differential operators refer to which
function they operate on, i.e. either ${\cal C} {\left( {\bf k}_1,
{\bf s}_1 \right)}$ or ${\cal C} {\left( {\bf k}_2, {\bf s}_2
\right)}$, never both. Bearing in mind that the functions ${\cal
A} {\left( {\bf k}_1, {\bf s}_1 \right)}$ and ${\cal B} {\left(
{\bf k}_2, {\bf s}_2 \right)}$ are independent of ${\bf x}$ and
${\bf p}$, we can take the differential operators outside the
integrals over ${\bf k}_1$, ${\bf s}_1$, ${\bf k}_2$ and ${\bf
s}_2$. As a result, we obtain
\begin{widetext}
\begin{equation}
\hbar \int {\rm d} {\bf x} {\rm d} {\bf p} D {\left( {\bf x}, {\bf
p} \right)} W {\left( {\bf x}, {\bf p}; t \right)} = 2 \int {\rm
d} {\bf x} {\rm d} {\bf p} W {\left( {\bf x}, {\bf p}; t \right)}
{\widehat{\sin}} {\frac {\hbar} {2}} {\left( \nabla_{xA} \cdot
\nabla_{pB} - \nabla_{pA} \cdot \nabla_{xB} \right)} A {\left(
{\bf x}, {\bf p} \right)} B {\left( {\bf x}, {\bf p} \right)},
\label{eq:identity3}
\end{equation}
\end{widetext}
where again the subscripts on the differential operators refer to
the corresponding function they operate on, i.e. either $A {\left(
{\bf x}, {\bf p} \right)}$ or $B {\left( {\bf x}, {\bf p}
\right)}$, never both. Comparing the non-Wigner function terms
inside the integrals immediately implies
\begin{equation}
{\frac {\hbar} {2}}  D {\left( {\bf x}, {\bf p} \right)} \nonumber
\end{equation}
\begin{equation}
= {\widehat{\sin}} {\frac {\hbar} {2}} {\left( \nabla_{xA} \cdot
\nabla_{pB} - \nabla_{pA} \cdot \nabla_{xB} \right)} A {\left(
{\bf x}, {\bf p} \right)} B {\left( {\bf x}, {\bf p} \right)},
\label{eq:moyalbra}
\end{equation}
which is the final result. Considering only the lowest order term
in $\hbar$, this reduces to
\begin{equation}
D {\left( {\bf x}, {\bf p} \right)} = {\frac {\partial A {\left(
{\bf x}, {\bf p} \right)}} {\partial {\bf x}}} \cdot {\frac
{\partial B {\left( {\bf x}, {\bf p} \right)}} {\partial {\bf p}}}
- {\frac {\partial A {\left( {\bf x}, {\bf p} \right)}} {\partial
{\bf p}}} \cdot {\frac {\partial B {\left( {\bf x}, {\bf p}
\right)}} {\partial {\bf x}}}. \label{eq:poissonbra}
\end{equation}

A common shorthand notation for equation (\ref{eq:moyalbra}) is
\begin{equation}
{\frac {i \hbar} {2}}  D {\left( {\bf x}, {\bf p} \right)} =
{\left\{ A {\left( {\bf x}, {\bf p} \right)}, B {\left( {\bf x},
{\bf p} \right)} \right\}}_{MB}, \label{eq:moyalbracket}
\end{equation}
where the subscript $MB$ stands for Moyal bracket, associating
this expression with both the Poisson bracket and the commutator
bracket.

\renewcommand{\theequation}{\thesection.\arabic{equation}}

\setcounter{equation}{0}

\section{Discussion and Conclusions}

Starting from a simple classical framework and employing some
stochastic concepts, the basic ingredients characterizing the
quantum nature of physical processes are recovered. It has been
shown that the traditional axiomatic structure of quantum
mechanics can be rebuilt, so that the quantum mechanical framework
resembles to a large extent that of the classical statistical
mechanics.

The main assumption used in the present paper is the existence of
a random irrotational component in the classical momentum. The
physical grounds for such assumption are that an ideal particle
localization is not feasible. Hence, provided a probability
density in configuration space is prescribed, the infinitesimal
variation of the particle "position" in the course of time (i.e.,
the particle velocity) cannot be determined precisely. Therefore,
there is always some uncertainty in the specification of particle
momentum, which should strongly depend on the degree of particle
delocalization in configuration space. The approach pursued here
is by no means an attempt to build a realistic model of the
underlying momentum fluctuations, however some hints concerning
their higher-order correlation properties are presented. In
particular, the current velocity fluctuations are shown to be
related to the turbulent fluctuations in the standard picture of
Reynolds turbulence. The latter represents an interesting and
promising guideline for further investigations.

As a result of the investigation performed, various basic elements
of the quantum formalism (calculation of expectation values, the
Heisenberg uncertainty principle, the correspondence principle)
are recovered by applying traditional techniques, borrowed from
classical statistical mechanics.

Finally, it is worthwhile to mention that the link between the
formalism used in the deformation quantization method and the
usual techniques of classical statistical mechanics appears quite
natural in our approach.



\appendix

\section{Analysis of the Compatibility Condition (\ref{eq:functionalw})}

To verify the validity of the compatibility condition
(\ref{eq:functionalw}), we expand its both sides in a power series
in the variable ${\bf s}$. To third order, we have
\begin{equation}
{\left \langle \exp {\left( {\frac {im} {\hbar}} {\bf s} \cdot
{\bf V} \right)} \right \rangle} = 1 + {\frac {i s_k} {\hbar}}
\partial_k {\cal S} - {\frac {s_k s_n} {2 \hbar^2}} \partial_k {\cal
S} \partial_n {\cal S} \nonumber
\end{equation}
\begin{equation}
- {\frac {m^2 s_k s_n} {2 \hbar^2}} {\cal C}_{kn} - {\frac {i m^3
s_k s_l s_n} {6 \hbar^3}} \nonumber
\end{equation}
\begin{equation}
\times {\left( v_k v_l v_n + v_k {\cal C}_{ln} + v_l {\cal C}_{nk}
+ v_n {\cal C}_{kl} + {\cal D}_{kln} \right)} + \dots,
\label{eq:lhsexpand}
\end{equation}
where
\begin{equation}
{\cal D}_{kln} = {\left \langle {\widetilde{V}}_k
{\widetilde{V}}_l {\widetilde{V}}_n \right \rangle},
\label{eq:thirdcorrel}
\end{equation}
is the yet unknown third order correlation function of the random
velocity field. Similarly, for the right-hand-side of equation
(\ref{eq:functionalw}), we obtain
\begin{equation}
\psi {\left( {\bf x} + {\frac {\bf s} {2}}; t \right)} \psi^{\ast}
{\left( {\bf x} - {\frac {\bf s} {2}}; t \right)} = \varrho +
{\frac {s_k} {2}} {\left( \psi^{\ast} \partial_k \psi - \psi
\partial_k \psi^{\ast} \right)} \nonumber
\end{equation}
\begin{equation}
+ {\frac {s_k s_n} {8}} {\left( \psi^{\ast} \partial_k \partial_n
\psi - \partial_k \psi^{\ast} \partial_n \psi - \partial_n
\psi^{\ast} \partial_k \psi^{\ast} + \psi \partial_k \partial_n
\psi^{\ast} \right)} \nonumber
\end{equation}
\begin{equation}
+ {\frac {s_k s_l s_n} {48}} {\left( \psi^{\ast}\partial_k
\partial_l \partial_n \psi - \psi \partial_k \partial_l \partial_n
\psi^{\ast} \right)} \nonumber
\end{equation}
\begin{equation}
+ {\frac {s_k s_l s_n} {48}} {\left[ \partial_k \psi \partial_l
\partial_n \psi^{\ast} - \partial_k \psi^{\ast} \partial_l
\partial_n \psi + {\rm Sym} (k,l,n) \right]}, \label{eq:thirdwigner}
\end{equation}
where ${\rm Sym} (k,l,n)$ denotes four additional terms obtained
by cyclic permutation of the indices $k$, $l$ and $n$.

Taking into account the explicit expression (\ref{eq:generform})
for the correlation tensor ${\cal C}_{kn}$ with ${\cal F} = 0$,
and the amplitude-phase representation (\ref{eq:debroglie}) of the
wave function, it is a simple matter to verify in a
straightforward manner that equation (\ref{eq:functionalw}) is
satisfied up to second order in the expansion variable ${\bf s}$.
To satisfy the third order however, one needs to specify the
third-order correlator of the random part of the current velocity.
We obtain the following expression
\begin{equation}
{\cal D}_{kln} = - {\frac {\hbar^2} {12 m^2}} {\left( \partial_l
\partial_n v_k + \partial_n \partial_k v_l + \partial_k \partial_l
v_n \right)}. \label{eq:thirdcorr}
\end{equation}

No attempt is made neither to interpret, nor to provide a physical
model (which beyond doubt should be nonlocal) of a possible
source, underlying the velocity (momentum) fluctuations. However,
from expression (\ref{eq:thirdcorr}), it is clear that these
fluctuations are far from being Gaussian. A further insight on the
relation between the current velocity fluctuations and the
standard picture of Reynolds turbulence is given in Appendix B.

\section{Turbulent Fluctuations and Compatibility Condition
(\ref{eq:functionalw})}

To show that the results from Appendix A are consistent with the
model presented in Section IV, we rewrite equations
(\ref{eq:continuity}) and (\ref{eq:momentum}) as
\begin{equation}
{\frac {\partial \varrho} {\partial t}} + \nabla_x \cdot {\left(
\varrho {\bf V} \right)} = 0, \label{eq:continuityB}
\end{equation}
\begin{equation}
{\frac {\partial {\bf V}} {\partial t}} + {\bf V} \cdot \nabla_x
{\bf V} = {\frac {\bf F} {m}}, \label{eq:momentumB}
\end{equation}
From the assumption that the density $\varrho$ does not fluctuate,
it follows that the fluctuating part of the continuity equation
(\ref{eq:continuityB}) reduces to
\begin{equation}
\nabla_x \cdot {\widetilde{\bf V}} = - {\widetilde{\bf V}} \cdot
\nabla_x R. \label{eq:fluctpart}
\end{equation}
Averaging the equation for momentum balance (\ref{eq:momentumB})
and taking into account relation (\ref{eq:fluctpart}), we readily
obtain equation (\ref{eq:momentu}).

Let us further write
\begin{equation}
{\frac {\partial} {\partial t}} {\left( V_n V_s \right)} + V_k
{\frac {\partial} {\partial x_k}} {\left( V_n V_s \right)} =
{\frac {1} {m}} {\left( F_n V_s + F_s V_n \right)}.
\label{eq:secondmom}
\end{equation}
which follows directly from the equation for momentum balance
(\ref{eq:momentumB}). Averaging the last equation
(\ref{eq:secondmom}) and taking into account equation
(\ref{eq:momentu}), we obtain
\begin{equation}
{\frac {\partial {\cal C}_{ns}} {\partial t}} + v_k {\frac
{\partial {\cal C}_{ns}} {\partial x_k}} + {\cal C}_{nk} {\frac
{\partial v_s} {\partial x_k}} + {\cal C}_{sk} {\frac {\partial
v_n} {\partial x_k}} + {\frac {\partial {\cal D}_{nsk}} {\partial
x_k}} + {\cal D}_{nsk} {\frac {\partial R} {\partial x_k}}
\nonumber
\end{equation}
\begin{equation}
= {\frac {1} {m}} {\left \langle V_s {\frac {\partial A_n}
{\partial t}} + V_n {\frac {\partial A_s} {\partial t}} \right
\rangle}. \label{eq:secmomaver}
\end{equation}
Having already determined the correlation tensor ${\cal C}_{ns}$
in the form (\ref{eq:generform}), we can manipulate the first term
on the left-hand-side of equation (\ref{eq:secmomaver}) using the
continuity equation (\ref{eq:continuit}). We again consider the
case, where ${\cal F} = 0$. The result is
\begin{equation}
{\cal C}_{sk} {\left( {\frac {\partial v_n} {\partial x_k}} -
{\frac {\partial v_k} {\partial x_n}} \right)} + {\cal C}_{nk}
{\left( {\frac {\partial v_s} {\partial x_k}} - {\frac {\partial
v_k} {\partial x_s}} \right)} \nonumber
\end{equation}
\begin{equation}
- \alpha {\left( {\frac {\partial^3 v_k} {\partial x_n \partial
x_s \partial x_k}} + {\frac {\partial^2 v_k} {\partial x_n
\partial x_s}} {\frac {\partial R} {\partial x_k}} \right)} +
{\frac {\partial {\cal D}_{nsk}} {\partial x_k}} \nonumber
\end{equation}
\begin{equation}
+ {\cal D}_{nsk} {\frac {\partial R} {\partial x_k}} = {\frac {1}
{m}} {\left \langle V_s {\frac {\partial A_n} {\partial t}} + V_n
{\frac {\partial A_s} {\partial t}} \right \rangle}.
\label{eq:secmomequat}
\end{equation}
If the current velocity ${\bf v}$ is irrotational as specified by
(\ref{eq:eikonal}), the first two terms on the left-hand-side of
equation (\ref{eq:secmomequat}) vanish. Assuming also that the
current velocity fluctuation ${\widetilde{\bf V}}$ is uncorrelated
with the random part of the force ${\bf F}$, we readily arrive
(after symmetrization) at the expression (\ref{eq:thirdcorr}) for
the triple correlation tensor ${\cal D}_{nsk}$.

Using equations (\ref{eq:momentumB}) and (\ref{eq:secondmom}) one
can proceed in calculation of the fourth-order correlator of the
current velocity fluctuation ${\widetilde{\bf V}}$. This procedure
can be continued further and all higher order correlators can be
found in principle.

\bibliography{apssamp}

\begin{thebibliography} {}

\bibitem{Weyl}
H. Weyl, Z. Phys., {\bf 40}, 1 (1927).

\bibitem{Wigner}
E. Wigner, Phys. Rev., {\bf 40}, 749 (1932).

\bibitem{Koopman}
B.O. Koopman, Proc. Natl. Acad. Sci. USA, {\bf 17}, 315 (1931).

\bibitem{Neumann1}
J. von Neumann, Ann. Math., {\bf 33}, 587 (1932).

\bibitem{Neumann2}
J. von Neumann, Ann. Math., {\bf 33}, 789 (1932).

\bibitem{Bohm}
D. Bohm, Phys. Rev., {\bf 85}, 166 (1952).

\bibitem{Vigier}
D. Bohm and J.P. Vigier, Phys. Rev., {\bf 96}, 208 (1954).

\bibitem{Fenyes}
I. F\'{e}nyes, Z. Phys., {\bf 132}, 81 (1952).

\bibitem{Nelson1}
E. Nelson, Phys. Rev., {\bf 150}, 1079 (1966).

\bibitem{Nelson2}
E. Nelson, {\it Dynamical Theories of Brownian Motion} (Princeton
University Press, Princeton, 1967).

\bibitem{Nelson3}
E. Nelson, {\it Quantum Fluctuations} (Princeton University Press,
Princeton, 1985).

\bibitem{Polavieja}
G.G. Polavieja, Phys. Lett. A, {\bf 220}, 303 (1996).

\bibitem{Leavens}
C.R. Leavens and R.S. Mayato, Phys. Lett. A, {\bf 280}, 163
(2001).

\bibitem{Prata1}
N.C. Dias and J.N. Prata, Phys. Lett. A, {\bf 291}, 355 (2001).

\bibitem{Prata2}
N.C. Dias and J.N. Prata, Phys. Lett. A, {\bf 302}, 261 (2002).

\bibitem{Hall}
M.J.W. Hall and M. Reginatto, J. Phys. A, {\bf 35}, 3289 (2002).

\bibitem{Reginatto}
M. Reginatto, Phys. Rev. A, {\bf 58}, 1775 (1998).

\bibitem{Tzenov}
S.I. Tzenov, {\it Random Beam Propagation in Accelerators and
Storage Rings}, FNT/T-95/27, University of Pavia, 1995,
physics/9908012.

\bibitem{Madelung}
E. Madelung, Z. Phys., {\bf 40}, 322 (1926).

\bibitem{Olavo}
L.S.F. Olavo, Physica A, {\bf 262}, 197 (1999).

\bibitem{Pesci}
A.I. Pesci and R.E. Goldstein, Nonlinearity, {\bf 18}, 211 (2005).

\bibitem{Kullback}
S. Kullback, {\it Information Theory and Statistics} (Wiley, New
York, 1959; corrected and revised edition, Dover Publications,
Inc., New York, 1968).

\bibitem{Birula}
I. Bialynicki-Birula and J. Mycielski , Annals of Physics, {\bf
100}, 62 (1976).

\bibitem{Moyal}
J. E. Moyal, Proc. Cambridge Phil. Soc. {\bf 45}, 99 (1949).

\bibitem{Gardiner}
C.W. Gardiner, {\it Quantum Noise} (Springer-Verlag, Berlin,
1991).





\end{thebibliography}

\end{document}